\documentclass{article}

\usepackage[final]{timeseries_workshop}

\usepackage[utf8]{inputenc} 
\usepackage[T1]{fontenc}    
\usepackage{hyperref}       
\usepackage{url}            
\usepackage{booktabs}       
\usepackage{amsfonts}       
\usepackage{nicefrac}       
\usepackage{microtype}      
\usepackage{xcolor}         
\usepackage{graphicx}
\usepackage{amsmath}
\usepackage{multirow}
\usepackage{graphicx}
\usepackage{subcaption}
\usepackage[export]{adjustbox}
\usepackage{wrapfig}

\title{Incorporating Metabolic Information into LLMs for Anomaly Detection in Clinical Time-Series}

\author{%
  Maxx Richard Rahman \\
  Saarland University \& DFKI\\
  Germany \\
  \texttt{m.rahman@iss.uni-saarland.de} \\
  \And
  Ruoxuan Liu \\
  DFKI\\
  Germany \\
  \texttt{ruoxuan.liu@dfki.de} \\
  \And
  Wolfgang Maass \\
  Saarland University \& DFKI \\
  Germany \\
  \texttt{wolfgang.maass@dfki.de} \\
}

\begin{document}

\maketitle

\begin{abstract}
  Anomaly detection in clinical time-series holds significant potential in identifying suspicious patterns in different biological parameters. This paper proposes a targeted method that incorporates the clinical domain knowledge into LLMs to improve their ability to detect anomalies. The Metabolism Pathway-driven Prompting (MPP) approach is introduced, which integrates the information about metabolic pathways to better capture the structural and temporal changes in biological samples. We applied our method for doping detection in sports, focusing on steroid metabolism, and evaluated using real-world data from athletes. The results show that our method improves anomaly detection performance by leveraging metabolic context, providing an improved prediction of suspicious samples in athletes' profiles.
\end{abstract}

\section{Introduction}

Clinical time series, also known as longitudinal profiles of individuals, represent repeated measurements of biological samples such as blood, urine, or other biological specimens collected over time [1,2]. These profiles are important in capturing the dynamic nature of biological processes, as they provide a time-evolving perspective of various physiological processes. The biomarkers measured within these samples often reflect underlying metabolic pathways [3].

\begin{wrapfigure}{r}{0.25\textwidth}
\includegraphics[width=\linewidth]{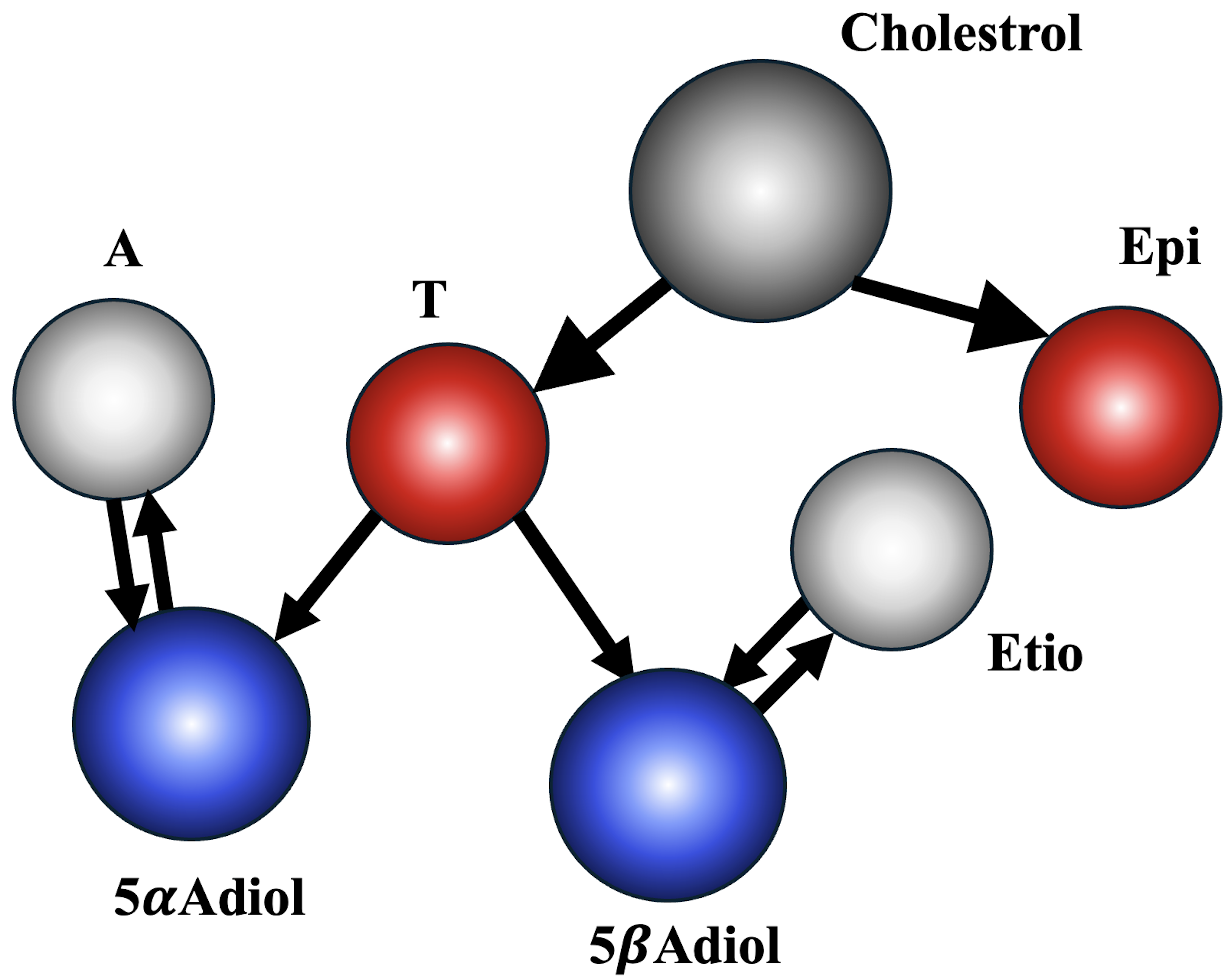} 
\caption{Simplified steroid pathway.
\label{fig:metabolism}}
\end{wrapfigure}

In clinical settings, anomaly detection in these longitudinal profiles is an important task [4]. Identifying abnormal behaviour in such data can reveal critical insights, ranging from disease diagnosis to sample tampering as potential doping activity in sports [5,6]. It mainly helps biochemists or clinicians to monitor biological and physiological changes over time and detect suspicious behaviour. Therefore, anomaly detection plays a significant role in both sports integrity and healthcare. Several studies have highlighted the potential and limitations of Large Language Models (LLMs) in clinical domain-specific tasks [7,8,9]. Despite their success in text generation, completion tasks, etc., their ability to process and analyse clinical time-series data, particularly in the context of metabolic pathways and biological changes, remains under explored [10,11]. Understanding how these models can leverage metabolic information to make informed decisions is critical for improving their performance in anomaly detection tasks.

This paper aims to understand the current capabilities of LLMs in handling longitudinal data and leveraging domain-specific knowledge for anomaly detection tasks. Specifically, a targeted prompting method is proposed by integrating metabolic pathway structures into LLMs to improve their ability to detect anomalies based on the contextual understanding. We demonstrate the effectiveness of our approach in doping detection in sports, where it is applied to detect suspicious urine samples within athletes' longitudinal profiles. These profiles include the concentrations of different metabolites, reflecting the steroid metabolism as shown in Fig.\ref{fig:metabolism}, and are important for identifying potential doping activities [12,13]. The key contributions of our paper can be summarised as follows:
\begin{itemize}
  \setlength\itemsep{0em}  
  \setlength{\leftskip}{-2em}  
  \item Metabolism Pathway-driven Prompting (MPP) is proposed, incorporating information about metabolic pathway structure and the temporal evolution of different metabolites into LLMs for anomaly detection task.
  \item  The effectiveness of this method is demonstrated in the context of doping detection in sports and compare it with the baseline prompting methods like zero-shot learning, in-context learning and chain-of-thought.
\end{itemize}

\section{Proposed Method}
\subsection{Problem Formulation}
Let the multivariate clinical time-series data be represented as longitudinal profile of athletes $X = \{ x_1, x_2, \dots, x_n \}$, where \( x_t \in \mathbb{R}^m \) represents the measurements at time \( t \), containing total \( m \) metabolite and \( x_{t,i} \) represents the measurement of metabolite \( i \) at time \( t \). The temporal difference is defined as \( \Delta x_{t,i}^T = x_{t,i} - x_{(t-1),i} \) representing the change in metabolite \( i \) over time. The anomaly detection task is to learn a function $f(x_t)$ that gives an anomaly score to each sample $x_t$ in the longitudinal profile $X$. The function flags the anomalous sample if the magnitude of the sum of \( \Delta x_{t,i}^T \) exceeds a predefined threshold \( \delta \), indicating significant deviation from the expected change:
\begin{equation}
f(x_t) = 
\begin{cases}
1, & \text{if } \Big|\sum_{i=1}^{m} \Delta x_{t,i}^T\Big| > \delta, \\
0, & \text{otherwise}.
\end{cases}
\end{equation}
The metabolic structural difference is defined as $\Delta x_{t,i}^M = x_{t,i} - x_{t,(i+1)}$ which needs to be considered.

\subsection{Metabolism Pathway-driven Prompting (MPP)}
\begin{figure}[hbt!]
\centering{\includegraphics[width=11cm]{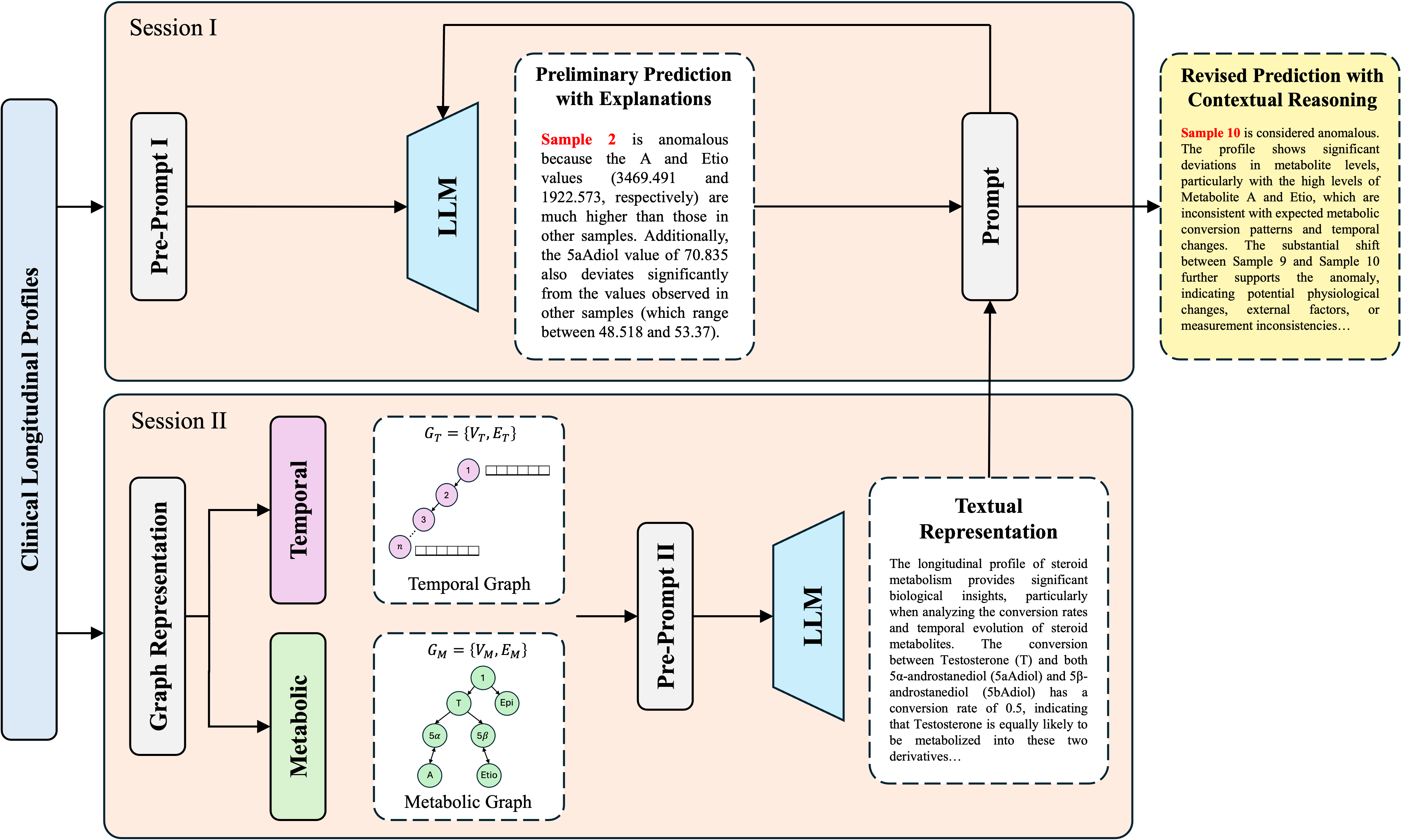}}
\caption{Schematic diagram of Metabolism Pathway-driven Prompting (MPP) method.
\label{fig:model}}
\end{figure}

A targeted prompting method is proposed, integrating metabolic pathway structures and their temporal evolution as shown in Fig.\ref{fig:model}. First, LLM (\textbf{Pre-Prompt I}) is tasked to analyse the longitudinal profile and detect anomalies using zero-shot learning. Here, the LLM usually considers the temporal changes between consecutive samples. If these changes exceed the statistically significant threshold, it flags the corresponding sample as anomalous with an explanation. In a different session, we input LLM (\textbf{Pre-Prompt II}) with temporal and metabolic graph representation of the given longitudinal profile and task to extract the domain-specific contextual information from these graph structures. The LLM generates a detailed textual explanation by assessing whether the temporal changes are consistent with the expected metabolic behaviour based on known pathways. Next, the textual representation of domain knowledge is provided to the previous LLM, which is then tasked to rethink (\textbf{Prompt}) by incorporating this domain-specific information. The LLM refines the initial prediction by combining the domain-specific information and provide more accurate, and contextually aware prediction.

\paragraph{Temporal Graph} 
The graph \( G_T = (V_T, E_T) \) represents the change in concentration levels of different steroids over time. Nodes are defined as $V_T = \{ x_1, x_2, \dots, x_n \}$, where each node \( x_t \) corresponds to the sample in the longitudinal profile and the node feature represents the measurements for the \( m \) steroids. The edges $E_T = \{ w_T(x_1 \to x_2), w_T(x_2 \to x_3), \dots, w_T(x_{n-1} \to x_{n}) \}$ represent transitions between nodes over time, connecting the samples between successive time points and the edge weights as the Euclidean distance between the steroid levels at two time points and normalized to the range [0, 1], incorporating the changes in all \( m \) steroids. For the edge connecting \( x_{n-1} \) and \( x_{n} \), the weight could be calculated as:
\begin{equation}
w_T(x_{n-1} \to x_{n}) = \sqrt{\sum_{i=1}^{m} (x_{n-1,i} - x_{n,i})^2}
\end{equation}

\paragraph{Metabolic Graph} 
The graph \( G_M = (V_M, E_M) \) represents the directional flow of \( m \) different metabolites (in this case steroids) in the pathway. Nodes are defined as $V_M = \{ S_1, S_2, \dots, S_m \}$, where each node \( S_i \) represents a steroid. The edges \( E_M \) represent the interactions or metabolic conversions between these steroids. The weight of an edge \( w_M(S_i \to S_j) \) represents the conversion rate from steroid \( S_i \) to steroid \( S_j \), where \( i, j = 1, 2, \dots, m \) and \( i \neq j \). If there is no conversion between two steroids, the corresponding entry is zero.
\begin{equation}
E_M = 
\begin{pmatrix}
0 & w_M(S_1 \to S_2) & w_M(S_1 \to S_3) & \dots & w_M(S_1 \to S_m) \\
w_M(S_2 \to S_1) & 0 & w_M(S_2 \to S_3) & \dots & w_M(S_2 \to S_m) \\
w_M(S_3 \to S_1) & w_M(S_3 \to S_2) & 0 & \dots & w_M(S_3 \to S_m) \\
\vdots & \vdots & \vdots & \ddots & \vdots \\
w_M(S_m \to S_1) & w_M(S_m \to S_2) & w_M(S_m \to S_3) & \dots & 0
\end{pmatrix}
\end{equation}

\section{Experiments}
\paragraph{Datasets}
Two real-world datasets (\textbf{Steroid-M} and \textbf{Steroid-F}) were used, consisting of longitudinal steroid profiles collected from male and female athletes, respectively [14,15]. The Steroid-M dataset contains 755 profiles with 4214 samples and Steroid-F dataset contains 375 profiles with 2307 samples. The data contains less than 20\% anomalous longitudinal profile.

\paragraph{Models and Metrics}
xperiments are conducted using different open-source LLMs: (i) LLaMa 2-7B [16], (ii) Mistral-7B [17], (iii) Falcon-7B [18], and (iv) GPT2 [19]. These models are selected due to their efficiency in providing quicker results, which is particularly suitable for the size of the dataset. The performance of the proposed method is compared with various baseline prompting methods, including Zero-Shot prompting (ZS) [20], In-Context Learning (ICL) [21], and Chain-of-Thought (CoT) [22], as well as two non-LLM-based models, IsoForest [23] and $\beta$-VAE [24]. Classification metrics such as accuracy, sensitivity, specificity, and F1-score are used for the anomaly detection task.

\section{Results}
\paragraph{Performance Comparison} 
Table\ref{results} shows that by incorporating domain-specific knowledge of metabolic pathways, MPP improves the LLMs' understanding of clinical data, leading to better performance. For the LLaMA 2-7B model, MPP achieves an accuracy of 71.4\% and an F1 score of 57.0\%, outperforming ZS's 65.2\% accuracy and 40.3\% F1 score on Steroid-M. Notably, MPP improves both sensitivity and specificity, which is important in clinical settings to balance correctly identifying actual anomalies while minimising false positives. In contrast, ICL and CoT generally underperform due to their lack of domain-specific guidance, i.e., ICL with GPT2 on Steroid-M yields only 28.2\% accuracy and a negligible 0.2\% F1 score. This underperformance highlights the importance of incorporating domain knowledge, as MPP does, to improve model performance for specialised tasks like clinical anomaly detection. 

\begin{figure}[hbt!]
\centering{\includegraphics[width=14cm]{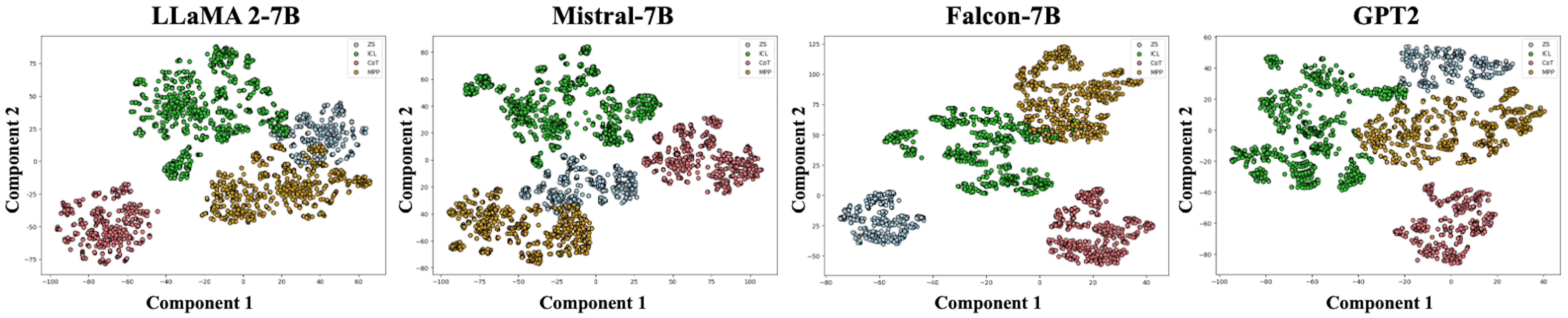}}
\caption{t-SNE representation of embeddings of the output from different prompting methods.
\label{fig:embeddings}}
\end{figure}

\paragraph{t-SNE Representation of Embeddings} 
Fig.\ref{fig:embeddings} shows the cluster formation in the embedding space of the LLM output which represents the distinct latent patterns captured by each prompting method. Across all models, the MPP forms well-defined clusters, indicating that it consistently produces more structured and distinct embeddings compared to the other prompting methods. This suggests that MPP effectively captures relevant patterns for anomaly detection in clinical data, outperforming the more dispersed clustering seen in ZS and ICL. Notably, CoT also produces structured clusters, but MPP shows greater distinction and compactness, especially in LLaMA 2-7B and Mistral-7B, highlighting the efficacy of pathway-driven prompting.

\begin{table}
\caption{Performance comparison of our proposed method with different baseline methods.}
\label{results}
\begin{center}
\begin{tabular}{cccccccccc}
\toprule
\multirow{2}{*}{\textbf{Model}} & \multirow{2}{*}{\textbf{Method}} & \multicolumn{4}{c}{\textbf{Steroid-M}}  &  \multicolumn{4}{c}{\textbf{Steroid-F}}  \\
\cline{3-10}
&  & \textbf{Acc.}  & \textbf{Sens.}  & \textbf{Spec.}  & \textbf{F1}  & \textbf{Acc.}  & \textbf{Sens.}  & \textbf{Spec.}  &
\textbf{F1}  \\
\midrule
\multirow{4}{*}{LLaMA 2-7B} & ZS & 0.652 & 0.912 & 0.563 & 0.403 & 0.402 & 0.567 & 0.382 & 0.250 \\

& ICL & 0.563 & 0.012 & 0.710 & 0.005 & 0.458 & 0.008 & 0.506 & 0.002 \\

& CoT & 0.228 & 0.526 & 0.130 & 0.208 & 0.426 & 0.506 & 0.381 & 0.250 \\

& \textbf{MPP} & \textbf{0.714} & \textbf{0.966} & 0.630 & \textbf{0.570} & 
\textbf{0.634} & \textbf{0.922} & 0.464 & \textbf{0.592} \\
\midrule
\multirow{4}{*}{Mistral-7B} & ZS & 0.763 & 0.931 & 0.632 & 0.578 & 0.724 & 0.012 & 0.905 & 0.028 \\

& ICL & 0.834 & 0.920 & 0.753 & 0.677 & 0.506 & 0.026 & 0.636 & 0.009 \\

& CoT & 0.501 & 0.894 & 0.598 & 0.517 & 0.626 & 0.012 & 0.752 & 0.002 \\

& \textbf{MPP} & \textbf{0.895} & 0.928 & \textbf{0.882} & \textbf{0.808} & \textbf{0.758} & \textbf{0.356} & 0.893 & \textbf{0.198} \\
\midrule
\multirow{4}{*}{Falcon-7B} & ZS & 0.352 & 0.960 & 0.125 & 0.364 & 0.395 & 0.308 & 0.474 & 0.406 \\

& ICL & 0.560 & 0.014 & 0.710 & 0.005 & 0.527 & 0.472 & 0.536 & 0.338 \\

& CoT & 0.524 & 0.673 & 0.432 & 0.440 & 0.388 & 0.024 & 0.383 & 0.008 \\

& \textbf{MPP} & \textbf{0.767} & 0.950 & \textbf{0.632} & \textbf{0.578} & \textbf{0.684} & \textbf{0.820} & 0.522 & \textbf{0.605} \\
\midrule
\multirow{4}{*}{GPT2} & ZS & 0.326 & 0.456 & 0.282 & 0.202 & 0.201 & 0.284 & 0.191 & 0.125 \\

& ICL & 0.282 & 0.006 & 0.355 & 0.002 & 0.229 &  0.004 & 0.253 & 0.001 \\

& CoT & 0.114 & 0.263 & 0.065 & 0.104 & 0.213 & 0.253 &  0.190 & 0.125 \\

& \textbf{MPP} & \textbf{0.357} & \textbf{0.483} & 0.315 & \textbf{0.285} & \textbf{0.317} & \textbf{0.461} & 0.232 & \textbf{0.296} \\

\midrule
\multirow{2}{*}{Non-LLM} & IsoForest & 0.786 & 0.296 & 0.985 & 0.451 & 0.719 &  0.364 & 0.986 & 0.528 \\

& $\beta$-VAE & 0.752 & 0.006 & \textbf{0.992} & 0.012 & 0.681 & 0.002 & \textbf{0.994} & 0.004 \\

\bottomrule
\end{tabular}
\end{center}
\end{table}

\section{Conclusion}
The Metabolism Pathway-driven Prompting (MPP) method is proposed to improve the anomaly detection within longitudinal data. By integrating the metabolic and temporal graphs for contextual understanding, MPP improves LLM's ability to detect anomalies, particularly in steroid metabolism, which is important for doping detection in sports. The results show improved accuracy and sensitivity compared to conventional prompting methods, demonstrating the significance of incorporating domain-specific knowledge for more precise and effective anomaly detection in clinical applications.

\section*{References}
{
\small
[1] Schüssler-Fiorenza Rose, S. M., Contrepois, K., Moneghetti, K. J., Zhou, W., Mishra, T., Mataraso, S., Snyder, M. P. (2019). A longitudinal big data approach for precision health. Nature medicine, 25(5), 792-804.

[2] Albert, P. S. (1999). Longitudinal data analysis (repeated measures) in clinical trials. Statistics in medicine, 18(13), 1707-1732.

[3] Monteiro, M. S., Carvalho, M., Bastos, M. L., Guedes de Pinho, P. (2013). Metabolomics analysis for biomarker discovery: advances and challenges. Current medicinal chemistry, 20(2), 257-271.

[4] Huang, Z., Lu, X., Duan, H. (2012). Anomaly detection in clinical processes. AMIA ... Annual Symposium proceedings. AMIA Symposium, 2012, 370–379.

[5] Moston, S., Engelberg, T. (2016). Detecting doping in sport. Routledge.

[6] Kumar, Y., Koul, A., Singla, R., Ijaz, M. F. (2023). Artificial intelligence in disease diagnosis: a systematic literature review, synthesizing framework and future research agenda. Journal of ambient intelligence and humanized computing, 14(7), 8459–8486. 

[7] Karabacak, M., Margetis, K. (2023). Embracing large language models for medical applications: opportunities and challenges. Cureus, 15(5).

[8] Nazi, Z. A., Peng, W. (2024, August). Large language models in healthcare and medical domain: A review. In Informatics (Vol. 11, No. 3, p. 57). MDPI.

[9] Neveditsin, N., Lingras, P., Mago, V. (2024). Clinical Insights: A Comprehensive Review of Language Models in Medicine. arXiv preprint arXiv:2408.11735.

[10] Helmy, M., Smith, D., Selvarajoo, K. (2020). Systems biology approaches integrated with artificial intelligence for optimized metabolic engineering. Metabolic engineering communications, 11, e00149.

[11] Hager, P., Jungmann, F., Bhagat, K., Hubrecht, I., Knauer, M., Vielhauer, J., Rueckert, D. (2024). Evaluating and Mitigating Limitations of Large Language Models in Clinical Decision Making. medRxiv, 2024-01.

[12] Piper, T., Geyer, H., Haenelt, N., Huelsemann, F., Schaenzer, W., Thevis, M. 2021. Current Insights into the Steroidal Module of the Athlete Biological Passport. International Journal of Sports and Medicine (42).

[13] Rahman, M. R., Hussain, M., Piper, T., Geyer, H., Equey, T., Baume, N., Aikin, R., Maass, W. (2023). Modelling Metabolism Pathways using Graph Representation Learning for Fraud Detection in Sports. In 2023 IEEE International Conference on Digital Health (ICDH) (pp. 158-168).

[14] Rahman, M.R., Piper, T., Geyer, H., Equey, T., Baume, N., Aikin, R., Maass, W. 2022. Data Analytics for Uncovering Fraudulent Behaviour in Elite Sports. In Proceedings of the 43rd International Conference on Information Systems (ICIS 2022).

[15] Rahman, M.R., Khaliq, L.A., Piper, T., Geyer, H., Equey, T., Baume, N., Aikin, R., Maass, W. (2024). SACNN: Self Attention-based Convolutional Neural Network for Fraudulent Behaviour Detection in Sports. In Proceedings of the International Joint Conference on Artificial Intelligence (IJCAI 24).

[16] Touvron, H., Martin, L., Stone, K., Albert, P., Almahairi, A., Babaei, Y., Scialom, T. (2023). Llama 2: Open foundation and fine-tuned chat models. arXiv preprint arXiv:2307.09288.

[17] Jiang, A. Q., Sablayrolles, A., Mensch, A., Bamford, C., Chaplot, D. S., Casas, D., Sayed, W. E. (2023). Mistral 7B. arXiv preprint arXiv:2310.06825.

[18] Almazrouei, E., Alobeidli, H., Alshamsi, A., Cappelli, A., Cojocaru, R., Debbah, M., Penedo, G. (2023). The falcon series of open language models. arXiv preprint arXiv:2311.16867.

[19] Radford, A., Wu, J., Child, R., Luan, D., Amodei, D., Sutskever, I. (2019). Language Models are Unsupervised Multitask Learners. OpenAI Blog (2019).

[20] Li, Y. (2023). A practical survey on zero-shot prompt design for in-context learning. arXiv:2309.13205.

[21] Yao, B., Chen, G., Zou, R., Lu, Y., Li, J., Zhang, S., Wang, D. (2024). More Samples or More Prompts? Exploring Effective Few-Shot In-Context Learning for LLMs with In-Context Sampling. In Findings of the Association for Computational Linguistics: NAACL 2024 (pp. 1772-1790).

[22] Cohn, C., Hutchins, N., Le, T., Biswas, G. (2024). A chain-of-thought prompting approach with llms for evaluating students’ formative assessment responses in science. In Proceedings of the AAAI Conference on Artificial Intelligence (Vol. 38, No. 21, pp. 23182-23190).

[23] Liu, F.T., Ting, K.M., and Zhou, Z.-H. (2008). Isolation-based anomaly detection. In 2008 Eighth IEEE International Conference on Data Mining, pages 413–422. IEEE, 2008.

[24] Higgins, I., Matthey, L., Pal, A., Burgess, C., Glorot, X., Botvinick, M., Mohamed, S., and Lerchner, A. (2017). beta-VAE: Learning basic visual concepts with a constrained variational framework. In International Conference on Learning Representations, 2017.

}



\pagebreak

\appendix

\section{Prompting Methods}

\subsection{Zero-Shot Learning} 
\begin{figure}[hbt!]
\centering{\includegraphics[width=12cm]{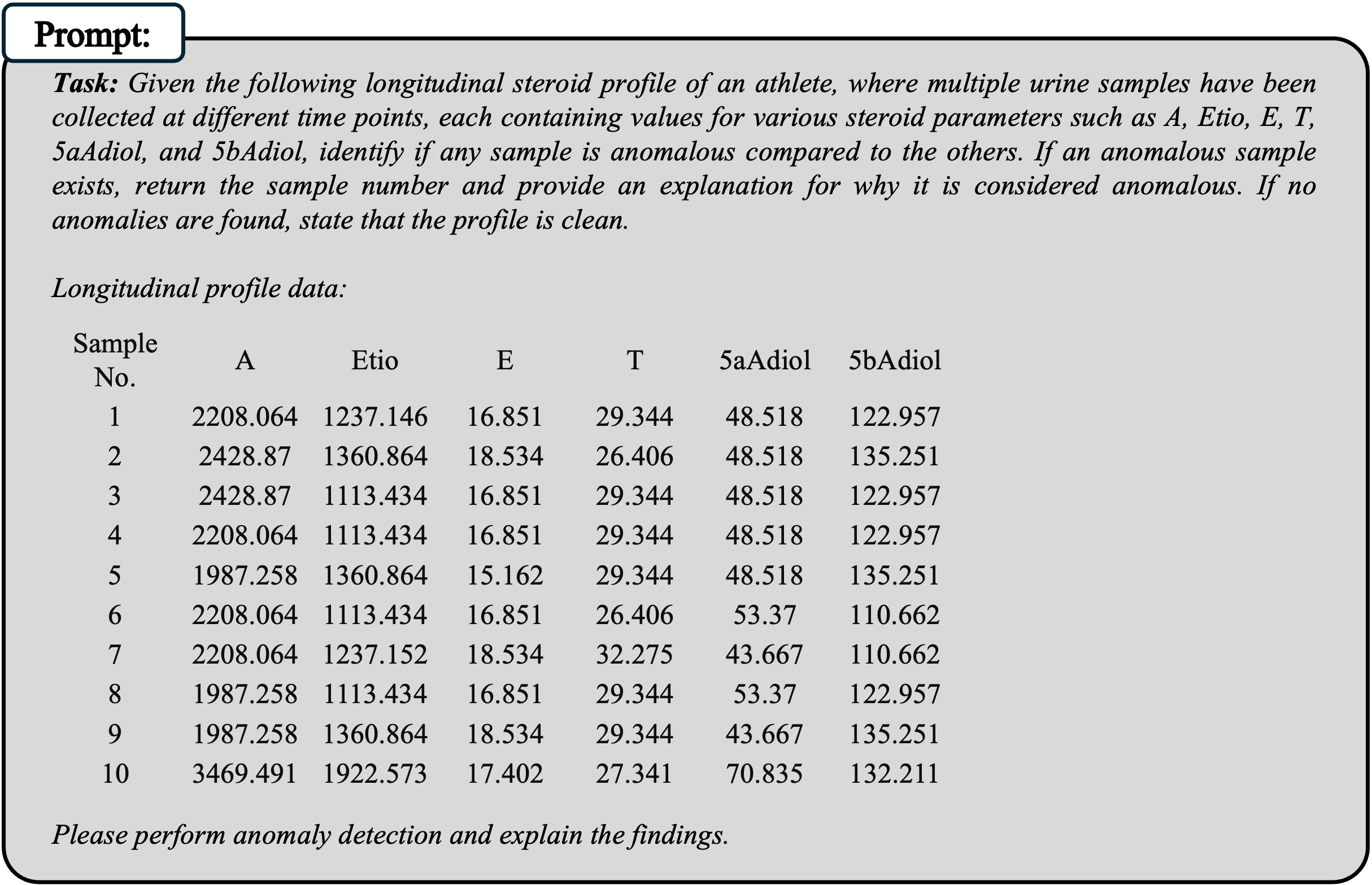}}
\caption{Example Prompt for Zero-shot Learning.
\label{fig:zero-shot}}
\end{figure}

It involves making predictions or identifying anomalies without explicit task-specific training. For longitudinal clinical data, we task LLM to infer whether certain samples are anomalous based on its prior general knowledge about statistical significance and typical clinical trends. Let $X = \{ x_1, x_2, \dots, x_n \}$, where \( x_t \in \mathbb{R}^m \) represents the measurements at time \( t \) and let \( A_t \) represent an anomaly indicator, where:
\begin{equation}
A_t =
\begin{cases} 
1, & \text{if an anomaly is detected at time } t, \\
0, & \text{otherwise}.
\end{cases}
\end{equation}
The model attempts to predict \( A_t \) directly from the input data \( X \) without prior training on labeled clinical anomaly data. Fig.\ref{fig:zero-shot} shows a prompt that we used for an example longitudinal profile.

\subsection{In-Context Learning} It involves providing the model with a few examples of what constitutes "normal" and "anomalous" patterns within the context of the prompt. The LLM uses these examples to generalise and apply its learned knowledge to unseen data points in the clinical data. Le us consider we are provided with \( k \) examples of clinical longitudinal profiles over time, which are labeled $\{ (X_1, A_1), (X_2, A_2), \dots, (X_k, A_k) \}$. These examples are included as part of the prompt. 

\begin{figure}[hbt!]
\centering{\includegraphics[width=12cm]{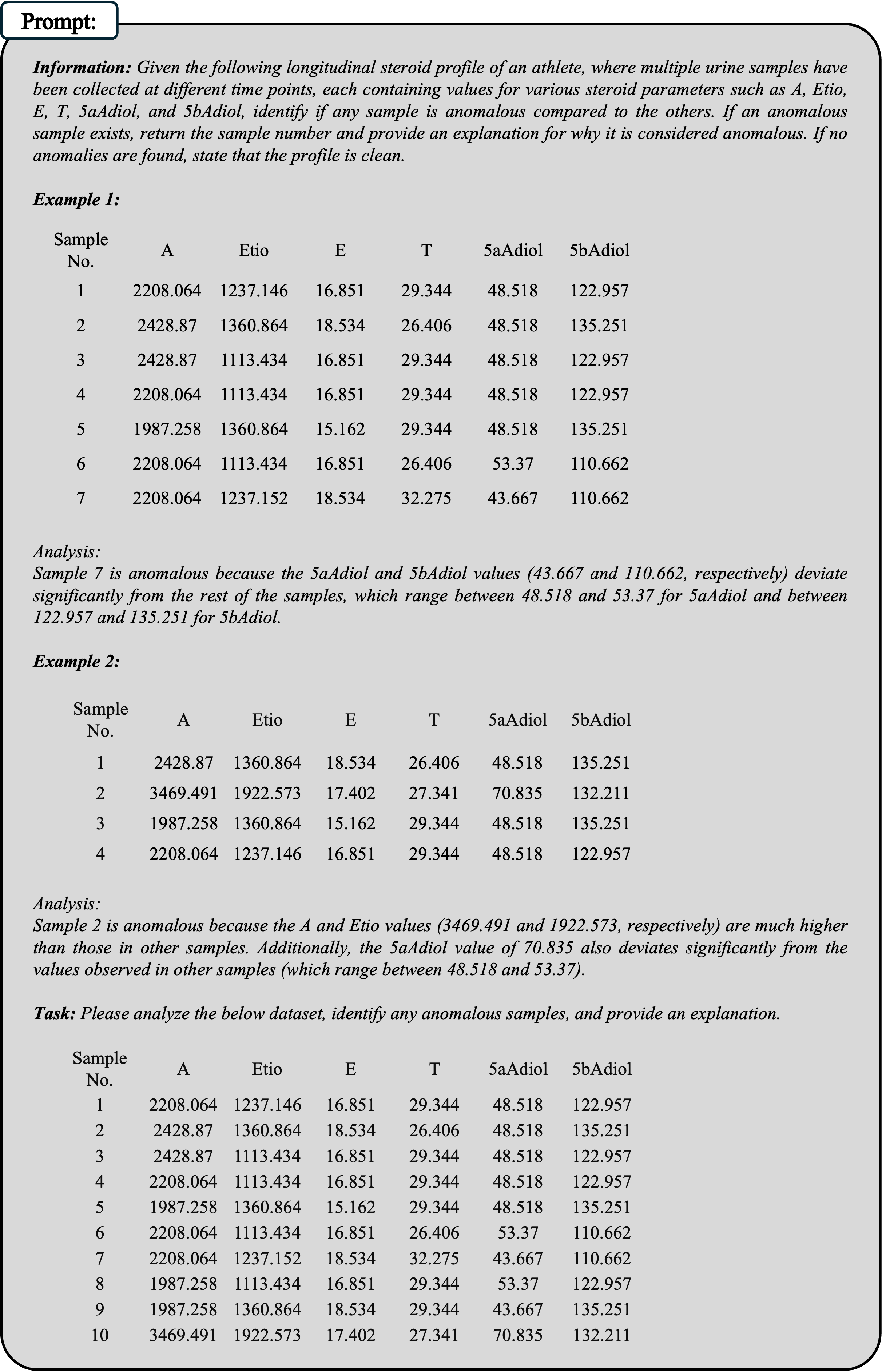}}
\caption{Example Prompt for In-context Learning.
\label{fig:icl}}
\end{figure}

Now, we give the longitudinal profile \( X \) to LLM to infer the anomaly label \( A_t \) by leveraging the information gained from the previous examples. 
\begin{equation}
A_t = f(X \mid \{ (X_1, A_1), (X_2, A_2), \dots, (X_k, A_k) \}),
\end{equation}
where function \( f(.) \) generalises from the given examples. Fig.\ref{fig:icl} shows a prompt that we used for an example longitudinal profile.

\subsection{Chain-of-Thought (CoT)} 
\begin{figure}[hbt!]
\centering{\includegraphics[width=12cm]{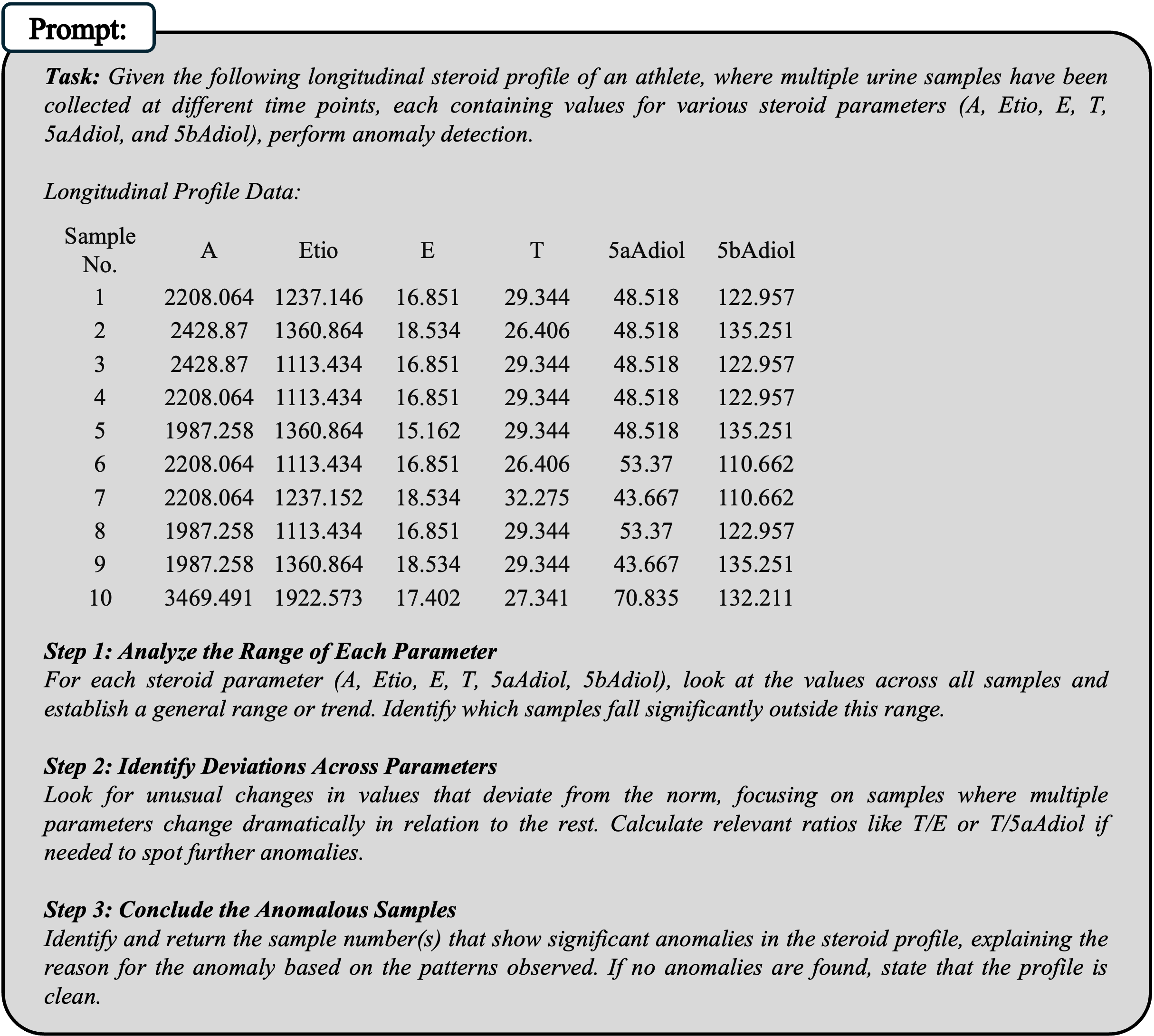}}
\caption{Example Prompt for Chain-of-Thought.
\label{fig:cot}}
\end{figure}

It encourages the model to reason through a multi-step process, explicitly following a logical progression before reaching its conclusion. For anomaly detection in longitudinal clinical data, this means that LLM is prompted to analyse how clinical values evolve over time and their relationships with other biomarkers before flagging an anomaly. We gave the following instructions:
\begin{itemize}
  \setlength\itemsep{0em}  
  \setlength{\leftskip}{-2em}  
  \item \textbf{Step 1: Analyse the range of each parameter} We task to examine the values of each steroid parameter across all the samples and calculate the mean (\(\mu\)) and standard deviation (\(\sigma\)) for each parameter to quantify the "normal" range:
  \begin{equation}
  \mu = \frac{1}{n} \sum_{i=1}^{n} x_i  \quad \text{and} \quad  \sigma = \sqrt{\frac{1}{n} \sum_{i=1}^{n} (x_i - \mu)^2}
  \end{equation}
  
  \item \textbf{Step 2: Identify deviations across parameters} After understanding the range of each parameter, we focus on finding deviations across multiple parameters within a sample to check if a particular sample shows unusually high or low values for more than one parameter. In addition, ratios between key parameters such as the Testosterone-to-Epitestosterone (T/E) ratio—can be useful indicators of abnormalities. Similarly, other useful ratios include: 
  \begin{equation}
  R_{T/E} = \frac{T}{E},  \hspace{0.7cm}  R_{T/5\alpha \, Adiol} = \frac{T}{5\alpha \, Adiol},  \hspace{0.7cm}  R_{T/5\beta \, Adiol} = \frac{T}{5\beta \, Adiol}
  \end{equation}
  Sudden deviations in these ratios compared to the rest of the samples can signal an abnormal sample.
 
  \item \textbf{Step 3: Conclude the anomalous samples} Finally, based on the findings from previous steps, we identify and return the sample number(s) that show significant anomalies. The anomaly is flagged based on either exceeding the calculated range (\(\mu \pm 2\sigma\)) or abnormal changes in the parameter ratios. If no significant anomalies are detected, the conclusion should state that the profile is clean. Fig.\ref{fig:cot} shows a prompt that we used for an example longitudinal profile.
\end{itemize}

\section{Metabolism Pathway-driven Prompting}
\begin{figure}[hbt!]
\centering{\includegraphics[width=12cm]{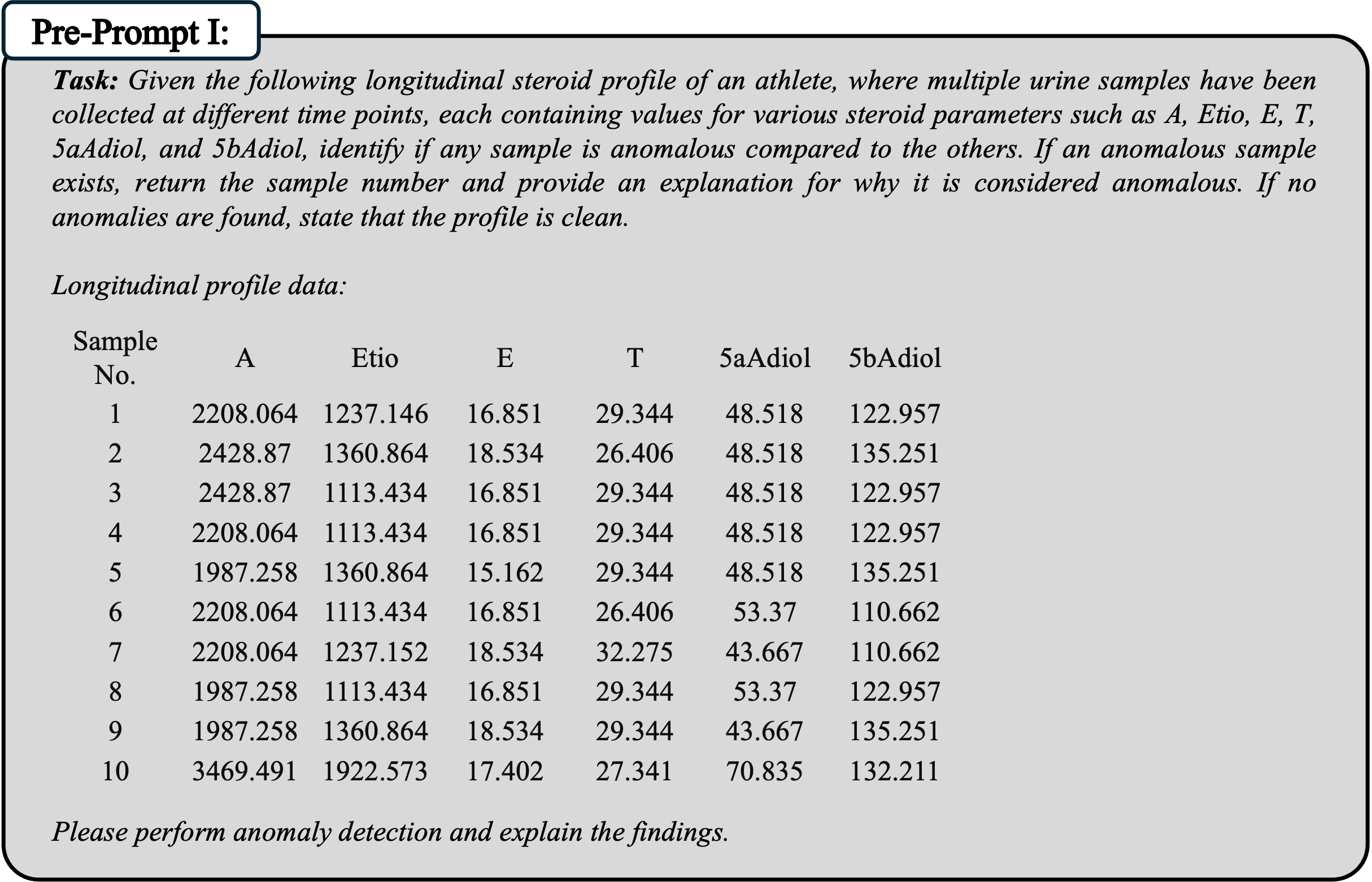}}
\caption{Pre-Prompt I for Metabolism Pathway-driven Prompting.
\label{fig:pre-promptI}}
\end{figure}

\begin{figure}[hbt!]
\centering{\includegraphics[width=12cm]{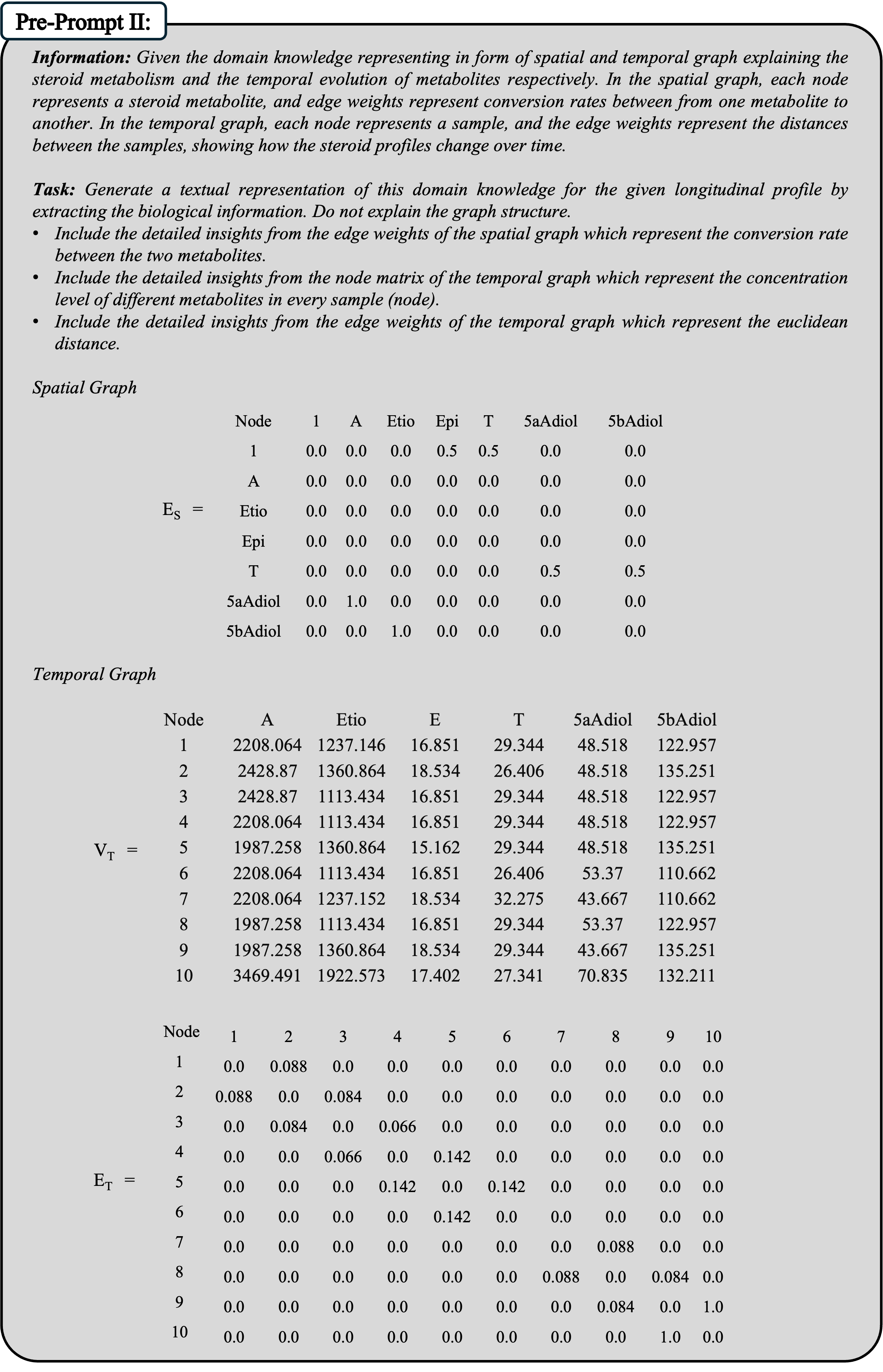}}
\caption{Pre-Prompt II for Metabolism Pathway-driven Prompting.
\label{fig:pre-promptII}}
\end{figure}

\begin{figure}[hbt!]
\centering{\includegraphics[width=12cm]{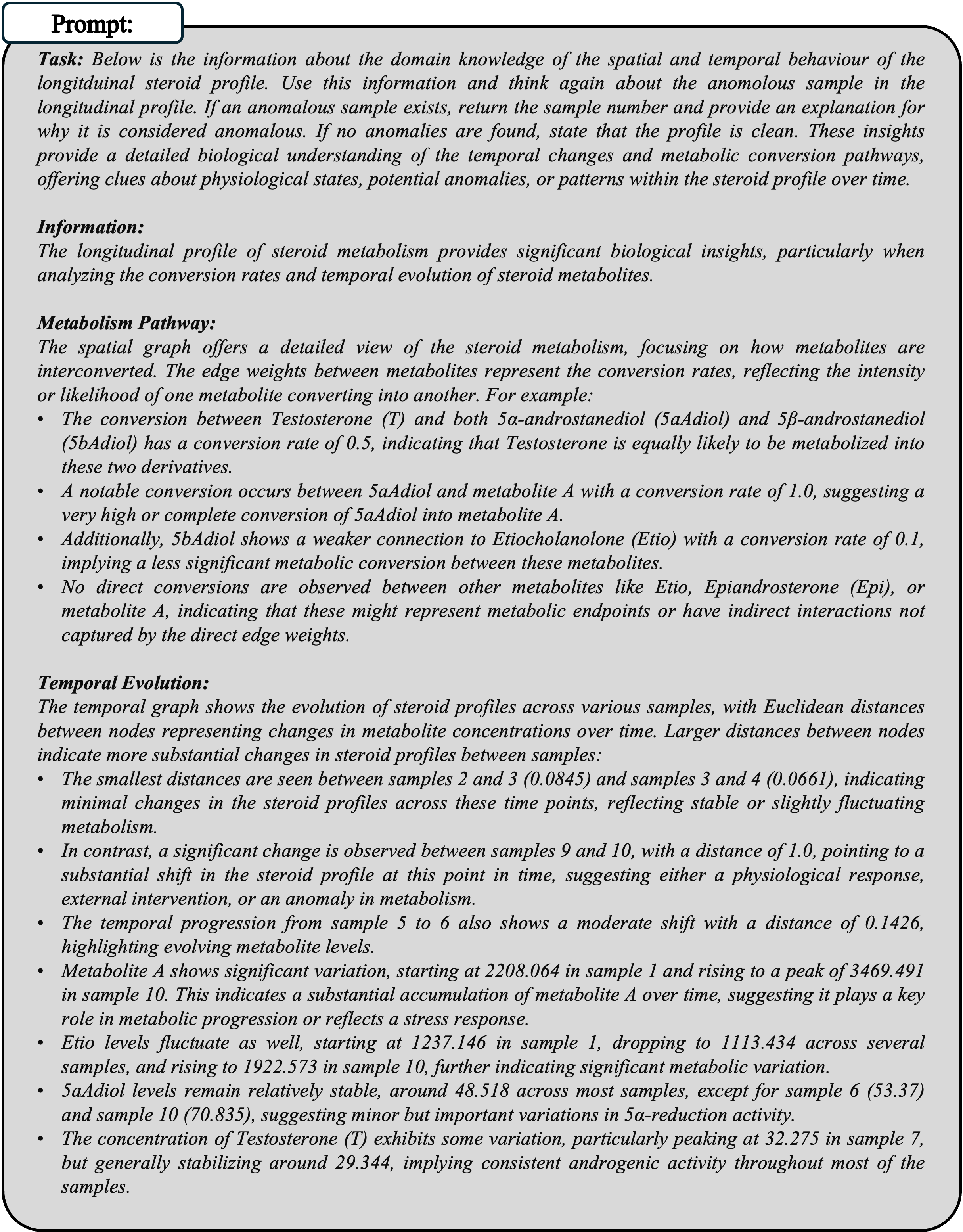}}
\caption{Prompt for Metabolism Pathway-driven Prompting.
\label{fig:prompt}}
\end{figure}

Fig.\ref{fig:pre-promptI} shows a Pre-prompt I for LLM that we used for an example longitudinal profile. Fig.\ref{fig:pre-promptII} shows a pre-prompt II that we used in another session for the same longitudinal profile. Finally, Fig.\ref{fig:prompt} shows a Prompt that integrates the response of the Pre-prompt II that we used for the same longitudinal profile.

\end{document}